\documentstyle[12pt]{article}

\begin{document}
\renewcommand{\thefootnote}{\fnsymbol{footnote}}
{\hfill hep-ph/9601244}
\vspace{1.0cm}
\begin{center}
{\huge \bf Sea polarisation and the Drell-Hearn-Gerasimov sum-rule(s) \\}
\vspace{3ex}
\vspace{3ex}
{\large \bf S. D. Bass \footnote{sbass@cernvm.cern.ch} \\}
\vspace{3ex}
{\it Institut f\"ur Kernphysik, KFA-J\"ulich, D-52425 J\"ulich, Germany \\ }
\vspace{3ex}

\vspace{3ex}

{\large \bf ABSTRACT \\}
\end{center}
\vspace{3ex}
{
The Drell-Hearn-Gerasimov sum-rule is really two sum-rules: one for each of the valence
and the sea/glue
contributions to the nucleon wavefunction.
The convergence of these sum-rules follows from the Froissart bound for spin dependent
processes in QCD
and is necessary for the consistency of the constituent quark model of low energy QCD.
Some challenges for future polarised photoproduction experiments, for example at ELFE,
are discussed.}

\newpage

\section{Introduction}

The EMC spin effect [1] has inspired a great deal of interest in QCD
spin physics ---
both at high and at low energies.
Polarised deep inelastic scattering
experiments are presently underway at CERN, DESY and SLAC to measure the spin dependent
nucleon structure functions 
at high energy and high $Q^2$ [2].
Experiments will soon begin at ELSA and MAMI [3] to measure the spin dependent
photoproduction cross sections 
with a view to testing the Drell-Hearn-Gerasimov sum-rule [4] (for reviews see [5,6]).
This paper is about the Drell-Hearn-Gerasimov sum-rule.
We explain why the no-subtraction hypothesis, which underlies the dispersion relation
derivation
of 
this sum-rule,
is correct because of the unitarity bound in QCD.
We then explain why the validity of this sum-rule is necessary for the constituent
quark model of
low energy QCD.
Finally, we discuss some challenges for future spin photoproduction experiments, 
for
example at the proposed future ELFE facility [7,8].

The Drell-Hearn-Gerasimov sum-rule [4,5] relates the difference in the spin dependent
inclusive
cross sections 
for a real transverse photon scattering from a longitudinally polarised
nucleon target
to the anomalous magnetic moment of the target nucleon.
Using standard
dispersion relation theory and assuming no need for subtractions, we consider
\begin{equation}
{\rm Re} f_2 (\nu) = {\nu \over 4 \pi^2}
\int_0^{\infty} d \nu' {\nu' \over \nu'^2 - \nu^2} 
( \sigma_{1 \over 2}^{\gamma N} -
  \sigma_{3 \over 2}^{\gamma N} )(\nu')
\end{equation}
where $\nu$ is the photon energy and
$f_2 (\nu)$ is the spin dependent part of the forward Compton scattering amplitude
\begin{equation}
f (\nu) = 
\chi_f^* \Biggl[
f_1 (\nu) {\vec \epsilon}_f^*.{\vec \epsilon}_i + i f_2 (\nu)
{\vec \sigma}.({\vec \epsilon}_f^* {\rm x} {\vec \epsilon}_i )
\biggr] \chi_i
\end{equation}
Photoproduction spin sum-rules are obtained by taking the odd derivatives (with
respect to $\nu$) of Re$f_2(\nu)$ in equ.(1),
viz.
$\biggl(
{\partial \over \partial \nu} \biggr)^{2n+1}$Re$f_2(\nu)$, and then evaluating
the resulting expression at $\nu =0$.
For small photon energy $\nu \rightarrow 0$,
\begin{equation}
f_1 (\nu) \rightarrow - {\alpha \over m} + ({\overline \alpha}_N + {\overline \beta}_N)
\nu^2
+ {\rm O}(\nu^4)
\end{equation}
and
\begin{equation}
f_2 (\nu) \rightarrow 
- {\alpha \kappa^2_N \over 2 m^2} \nu + \gamma_N \nu^3 + {\rm O}(\nu^5)
\end{equation}
Here $\kappa_N$ is the anomalous magnetic moment of the target nucleon, 
$m$ is
the nucleon mass,
and $\gamma_N$ and $({\overline \alpha}_N + {\overline \beta}_N)$ are
the charge parity $C=+1$
spin and
Rayleigh polarisabilities of the nucleon respectively.
The Drell-Hearn-Gerasimov sum-rule is derived when we evaluate the first derivative
($n=0$) of Re$f_2(\nu)$ in equ.(1) at $\nu =0$, viz.
\begin{equation}
\int_0^{\infty} {d \nu' \over \nu'} 
\Biggl ( \sigma_{1 \over 2}^{\gamma N} - \sigma_{3 \over 2}^{\gamma N} \Biggr )(\nu')
= - { 2 \pi^2 \alpha \over m^2 } \kappa^2_N
\end{equation}
The third derivative of Re$f_2 (\nu)$ yields a second sum-rule for the spin 
polarisibility $\gamma_N$
\begin{equation}
\int_0^{\infty} {d \nu' \over \nu'^3}
\Biggl (\sigma_{1 \over 2}^{\gamma N} - \sigma_{3 \over 2}^{\gamma N} \Biggr)(\nu')
=
{1 \over 4 \pi^2} \gamma_N
\end{equation}
These photoproduction sum-rules have a parallel in high $Q^2$ deep inelastic 
scattering where the odd moments
of the spin dependent structure function $g_1$
project out the nucleon matrix elements of gauge-invariant local operators, each of
which is multiplied
by a radiative coefficient.

\section {The validity of the no-subtraction hypothesis }

Besides the no-subtraction hypothesis,
the Drell-Hearn-Gerasimov 
sum-rule is derived assuming only very general principles:
electromagnetic gauge invariance, Lorentz invariance, causality and unitarity.
The no-subtraction hypothesis is the only ``QCD input" to the sum-rule. 
It is correct
because of the Froissart unitarity bound
for $\Delta \sigma = (\sigma_{1 \over 2}^{\gamma N} - \sigma_{3 \over 2}^{\gamma N})$
at large $\nu \rightarrow \infty$.
We consider the Regge theory [9], which works for both real and also complex
energy.
It is well known [10,11] from spin dependent Regge theory that the iso-triplet
contribution to
$\Delta \sigma$
behaves as
\begin{equation}
(\Delta \sigma)_3 (\nu) \sim \nu^{\alpha_{a_1} - 1}
\end{equation}
at large $\nu$.
Here $\alpha_{a_1}$ is the 
intercept
of the $a_1$ Regge trajectory ($-0.5 \leq \alpha_{a_1} \leq 0$ [12]) .
The iso-singlet contribution is still somewhat controversial and depends on
the spin
structure
of the short range part of the exchange potential [13].
The soft pomeron exchange, which dominates the large $\nu$ 
(centre of mass energy greater than about 10GeV) behaviour of the spin averaged 
cross section,
does not contribute to $\Delta \sigma$ [11].
However,
the physics which gives us the soft pomeron can contribute to $\Delta \sigma$ at
large photon energy $\nu$.
If the pomeron were a scalar, then we would find
\begin{equation}
(\Delta \sigma)_0^{[1]} \sim 0
\end{equation}
at large $\nu$.
In the Landshoff-Nachtmann approach [14], the soft pomeron is modelled
by the exchange of two non-perturbative
gluons
and behaves as a $C=+1$ vector potential with a correlation length of about 0.1fm.
This vector pomeron
gives
a contribution [13,15]
\begin{equation}
(\Delta \sigma)_0^{[2]} (\nu) \sim { \ln \nu \over \nu}
\end{equation}
to $\Delta \sigma$ at large $\nu$.
It has been suggested [16] that there may be a negative signature
two pomeron
cut contribution to $\Delta \sigma$ at large energy, viz.
\begin{equation}
(\Delta \sigma)_0^{[3]} (\nu) \sim { 1 \over \ln^2 \nu }
\end{equation}
although the coefficient of this term in the Regge calculus was found to be zero
in a recent re-analysis of the theory [17].
It is an important challenge to determine the strength of the three
possible
large $\nu$ contributions equs.(7-10)
in future photoproduction experiments and make contact with the small $x$
behaviour of the high $Q^2$ deep inelastic spin structure function $g_1$ [13,15,18]
which is also determined by Regge theory.
Each of the possible contributions in equs.(7-10)
give a convergent integral in equs.(1,5)
and support the
no-subtraction hypothesis that went into the derivation of the sum-rule.
The Regge theory works for all complex energy [9] so one finds a vanishing
contribution to the Drell-Hearn-Gerasimov sum-rule from ``the half circle"
at infinite complex momentum
which appears
when we close the contour in the dispersion relation derivation of the sum-rule.
We would have needed a subtraction if the sea were generated all with the same
polarisation,
which is not what happens in QCD.
(In this case
$\Delta \sigma$ would grow to reach the Froissart bound for the spin averaged
cross section
($\sigma_T \sim \ln^2 \nu$) and the integral in equs.(1,5) would not converge.)
Henceforth, we accept 
the Drell-Hearn-Gerasimov sum-rule as valid.

\section {The Drell-Hearn-Gerasimov sum-rule and the spin structure of the nucleon}

There is a beautiful property of the Drell-Hearn-Gerasimov sum-rule that has not
been
noted previously.
The inclusive spin dependent cross sections receive contributions from the photon's
coupling
to sea quarks and gluonic degrees of freedom as well as to the valence quarks.
At the same time, the anomalous magnetic moment $\kappa_N$ is a $C=-1$ 
(pure valence) quantity.
(It is measured in the nucleon's matrix element of the conserved, $C=-1$ vector 
current.)
The constituent quark model expression for the magnetic moment 
$\mu_N = \kappa_N + 1$
is:
\begin{equation}
\mu_N = \sum_q e_q {\sigma_q \over 2m_q}
\end{equation}
Here $e_q$ is the quark charge, $m_q$ is the constituent quark mass and
\begin{equation}
\sigma_q = (q - {\overline q})^{\uparrow} - (q - {\overline q})^{\downarrow}
\end{equation}
is the fraction of the nucleon's spin that is carried by the valence
quarks;
$q^{\uparrow (\downarrow)}$ denotes the probability for finding
a quark in the nucleon with flavour $q$ and 
polarised in the same (opposite) direction as 
the nucleon's polarisation; 
similarly ${\overline q}^{\uparrow (\downarrow)}$ for the anti-quark.
The valence spin content
$\sigma_q$ is invariant under $Q^2$ evolution. (It is $C$ odd and satisfies
the same $Q^2$ evolution equation as the non singlet axial charge $g_A^3$.)
Therefore,
it describes
equally the constituent quarks of low energy QCD and also the valence 
current quarks 
which are measured
in high $Q^2$ deep inelastic scattering.
The fact that $\kappa_N$ is a $C=-1$ pure valence quantity means that
one does not need to know anything about the Dirac sea or about 
the gluonic 
components
in the nucleon wavefunction in order to evaluate equ.(5).
The inverse-energy weighted integral of the glue and Dirac sea contributions to the
difference in 
the spin dependent cross sections at photoproduction is exactly zero.
The vanishing of
the polarised sea and gluonic
contribution to the Drell-Hearn-Gerasimov integral, equ.(5),
follows from the $C-$parity of the terms in the sum-rule and is derived assuming
only that the no subtraction hypothesis is correct.
In other words,
the general principles of
electromagnetic gauge invariance, Lorentz invariance, causality and unitarity
imply
that the polarised sea and gluonic contribution to the integral is either zero 
or
infinite
(so that one would need a subtraction).
In the infinite scenario
the sea and glue would be generated all with one polarisation, which is not QCD.
The vanishing sea and gluonic contribution to the Drell-Hearn-Gerasimov sum-rule
implies
an exact cancellation between the spin structure of the long and short range
parts of the $C= +1$
exchange potential.
If we introduce a cut-off $\nu_0$, then
\begin{equation}
\int_0^{\nu_0} 
{d \nu^{'} \over \nu^{'}} \Delta \sigma|_{\rm sea + glue} (\nu^{'})
= - 
\int_{\nu_0}^{\infty} 
{d \nu^{'} \over \nu^{'}} \Delta \sigma|_{\rm sea + glue} (\nu^{'})
\end{equation}
independent of the choice of cut-off $\nu_0$.
The long range and short range parts of the $C=+1$ exchange potential, which are
described by
chiral perturbation theory [19] and
the Bonn [20] and Paris [21] potentials of nuclear physics and by Regge theory
[10-17] respectively, 
``know about each other"
as, indeed,
they must in a consistent field theoretical description of the structure of the nucleon.

The pure valence nature of the Drell-Hearn-Gerasimov sum-rule allows us to
understand why the sum-rule is nearly saturated  
by the $N - \Delta$ transition [22,6], which is the leading valence quark spin
excitation.
It will be interesting to compare the relative saturation of the 
Drell-Hearn-Gerasimov and spin
polarisability sum-rules, which are each given by the difference of two positive
cross sections,
in future photoproduction experiments.
This requires first a direct and independent
measurement of $\gamma_N$ in a forward Compton scattering
experiment.
The experimental question to be tested is at what energy $\nu$ do we obtain 80\%, 90\% 
of the each of the two sum-rules ?
The spin polarisability sum-rule, equ.(6), receives contributions from the photon's
coupling to 
polarised sea and
gluonic excitations, as well as to the valence quarks, in the nucleon's wavefunction.

The validity of the high energy no-subtraction hypothesis is also necessary for
the
spin structure of the constituent quark model at low momentum scales.
The difference in the two spin cross sections can be written in terms of two spin
dependent
nucleon form factors $G_1(\nu, Q^2)$ and $G_2(\nu, Q^2)$, viz.
\begin{equation}
\sigma_{1 \over 2}^{\gamma N} - \sigma_{3 \over 2}^{\gamma N}
=
{16 m \pi^2 \alpha \over 2 m \nu - Q^2} 
\Biggl( 
m \nu G_1 (\nu, Q^2) - Q^2 G_2 (\nu, Q^2) 
\Biggr)
\end{equation}
Following Anselmino et al.[23], we define the $Q^2$ dependent quantity
\begin{equation}
I (Q^2) = m^3 \int_{ {Q^2 \over 2m} }^{\infty} {d \nu \over \nu} G_1 (\nu, Q^2)
\end{equation}
The Drell-Hearn-Gerasimov sum-rule tell us that
\begin{equation}
I(0) = - {1 \over 4} \kappa^2_N
\end{equation}
At the high $Q^2$ of deep inelastic scattering experiments
\begin{equation}
m^2 \nu G_1 (\nu, Q^2) \rightarrow g_1 (x, Q^2)
\end{equation}
so that
\begin{equation}
I (Q^2) =
{2 m^2 \over Q^2}
\int_0^1 dx g_1 (x, Q^2).
\end{equation}
The first moment of $g_1$ is
\begin{equation}
\int_0^1 dx g_1 (x, Q^2) = 
{1 \over 2} \sum_q e^2_q \Delta q (Q^2) C_q (1, \alpha_s(Q^2))
\end{equation}
Here
\begin{equation}
\Delta q (Q^2)
= (q + {\overline q})^{\uparrow}(Q^2) - (q + {\overline q})^{\downarrow}(Q^2)
\end{equation}
is the contribution to the nucleon's spin from quarks and anti-quarks of flavour $q$
and $C_q(1, \alpha_s(Q^2))$ is the radiative QCD coefficient.
(These $\Delta q$
include the effect of the axial anomaly [24].)
We compare equ.(20) with the conserved $C=-1$
spin coupling in equ.(12) which is measured in the Drell-Hearn-Gerasimov sum-rule.
The ${\overline q}$ sea polarisation contribution (which includes polarised glue
via the anomaly)
completely disappears at $Q^2=0$,
where higher twist effects have built up an effective spin operator for low momentum
scales.
The $C=-1$ spin coupling to $(q - {\overline q})$ measures a conserved quantity and
so
corresponds to the same spin operator for both current and constituent quarks.
We find an a posteriori justification of the spin structure of the constituent quark
model.

It is interesting to compare the physics of Drell-Hearn-Gerasimov with the spin 
averaged
cross section.
Here the no-subtraction hypothesis is not valid.
If we try to use an unsubtracted dispersion relation in the spin averaged case,
then we find an equation relating the total $\gamma N$ cross section to the Thomson 
term
in $f_1$ (equ.(3)),
which is given by
the
square of the forward matrix element of the $C=-1$ vector current,
which itself measures the
number of valence quarks in the nucleon.
If an unsubtracted dispersion relation were valid for the spin averaged cross-section
then the nucleon would have to be elementary!
Pomeron exchange, which tells us that total cross section grows as $\ln^2 \nu$ 
at very large
$\nu$, implies that the number of sea quarks is infinite in high $Q^2$ deep inelastic
scattering.
At low momentum scales the higher twist contributions become important and build up
effective constituent quark operators.
The infinite number of anti-quarks (and gluons) which are seen at high $Q^2$, and which
are manifest
as the need for a subtraction in the dispersion relation for spin averaged
photoproduction,
form a condensate in the vacuum at some low scale leading to spontaneous symmetry 
breaking
of chiral symmetry and pion physics.
The formation of this condensate is responsible for the large constituent quark mass.
Thus, photoproduction sum-rules
illustrate the self-consistency of our understanding of the structure of the nucleon
at low and
at high energies.

\section {Outlook}

Finally, we consider the challenges for future experiments.
The twin experiments at
MAMI and ELSA will measure the spin dependent photoproduction cross sections up
to a
photon energy of 3GeV [3].
They will cover the resonance region and the domain where we expect Regge theory to
start to work.
(Regge theory provides a good description of the unpolarised $\gamma p$ cross section
starting at
$\sqrt{s} \sim 2.5$GeV [25].)
It will be important to extend these polarised photoproduction experiments to higher
energies
to extract the relative strengths of the iso-triplet and
iso-singlet
Regge exchange contributions in equs.(7-10).
One could then make contact with the small $x$ behaviour of the spin structure function
$g_1$
of polarised deep inelastic scattering at high $Q^2$ -- see [13,15,18] 
for theory and
[26] for the latest data from SMC.
ELFE [7,8] will provide polarised real photon beams with $\sqrt {s} \simeq 7$GeV
(well within the domain of Regge physics); 
a highly polarised proton beam at HERA energy would take us deep into the region
of ``pomeron physics".
Secondly, it will be interesting to compare the relative saturation of
the (in practice $C=-1$) Drell-Hearn-Gerasimov and $C=+1$ spin polarisation sum-rules
in equs.(5,6)
to test
the spin structure of sea and gluonic excitations at photoproduction.

\vspace{1.0cm}
{\large \bf Acknowledgements: \\}
\vspace{3ex}

It is a pleasure to thank N. d'Hose, P.V. Landshoff, D. Sch\"utte and 
A.W. Thomas
for helpful
discussions.
This work was supported in part by the Alexander von Humboldt Foundation.

\pagebreak

\end{document}